# A search for thermal gyro-synchrotron emission from hot stellar coronae

Walter W. Golay[⬤],[1]★ Robert L. Mutel[⬤],[1]★ Dani Lipman[1,2] and Manuel Güdel[3]

[1]*Department of Physics and Astronomy, University of Iowa, Iowa City, IA 52242, USA*
[2]*Department of Physics, University of Connecticut, 196A Auditorium Road, Unit 3046, Storrs, CT 06269, USA*
[3]*Department of Astrophysics, University of Vienna, Türkenschanzstrasse 17 A-1180 Vienna, Austria*

## ABSTRACT

We searched for thermal gyro-synchrotron radio emission from a sample of five radio-loud stars whose X-ray coronae contain a hot ($T_e > 10^7$ K) thermal component. We used the JVLA to measure Stokes I and V/I spectral energy distributions (SEDs) over the frequency range 15 – 45 GHz, determining the best-fitting model parameters using power-law and thermal gyro-synchrotron emission models. The SEDs of the three chromospherically active binaries (Algol, UX Arietis, HR 1099) were well-fit by a power-law gyro-synchrotron model, with no evidence for a thermal component. However, the SEDs of the two weak-lined T Tauri stars (V410 Tau, HD 283572) had a circularly polarized enhancement above 30 GHz that was inconsistent with a pure power-law distribution. These spectra were well-fit by summing the emission from an extended coronal volume of power-law gyro-synchrotron emission and a smaller region with thermal plasma and a much stronger magnetic field emitting thermal gyro-synchrotron radiation. We used Bayesian inference to estimate the physical plasma parameters of the emission regions (characteristic size, electron density, temperature, power-law index, and magnetic field strength and direction) using independently measured radio sizes, X-ray luminosities, and magnetic field strengths as priors, where available. The derived parameters were well-constrained but somewhat degenerate. The power-law and thermal volumes in the pre-main-sequence stars are probably not co-spatial, and we speculate they may arise from two distinct regions: a tangled-field magnetosphere where reconnection occurs and a recently discovered low-latitude poloidal magnetic field, respectively.

## 1 INTRODUCTION

Non-thermal radio emission from stars is a powerful diagnostic for investigating high-energy processes in stellar coronae. In particular, gyro-synchrotron and synchrotron radiation from high-energy electrons spiraling in strong coronal magnetic fields is a common feature of many stellar systems ranging from 'normal' stars like the Sun (Bastian, Benz & Gary 1998) to evolved pre-main-sequence stars (Launhardt et al. 2022), close late-type binaries (Drake, Simon & Linsky 1992), and even ultra-cool dwarfs near the bottom of the main-sequence (Williams 2018).

The energetic electrons typically follow a power-law energy distribution, presumably accelerated by reconnection of magnetic fields in the corona. The accelerated electrons have energies with Lorentz factors of order 10 – 100, that is, mildly relativistic, so the radiation is referred to as gyro-synchrotron (GS) emission. This mechanism has been well-studied in a range of stellar environments. The observed spectral energy distributions (SEDs) and polarization have allowed quantitative estimates of the magnetic field strength and energetic particle densities in the solar and stellar coronae (e.g. Umana et al. 1993; Storey & Hewitt 1995; Mutel et al. 1998; Trigilio et al. 2001; García-Sánchez, Paredes & Ribó 2003; Leto et al. 2006;

Osten & Bastian 2006; Waterfall et al. 2019; Launhardt et al. 2022; Tan 2022).

A less well-known emission mechanism is gyro radiation from *thermal* electrons spiraling in a stellar magnetic field. If the electrons are non-relativistic, the radiation is called gyro-resonance, and emission (and absorption) occurs at the local electron gyro frequency, and its first few harmonics. Solar flares have demonstrated gyro-resonance emission (e.g. Nindos et al. 2000), often at the third harmonic.

For very hot plasmas ($T_e \gtrsim 10^7$ K), there are a significant number of electrons in the high-energy tail of a Maxwellian energy distribution, and since they are mildly relativistic, they radiate at low harmonics of the gyro-frequency. The resulting radiation–which can be highly circularly polarized depending on the aspect angle– is termed thermal GS radiation (Dulk 1985). Since thermal GS emission strongly depends on the magnetic field strength and coronal temperature, its detection can provide a sensitive measure of these quantities independent of assumptions about a power-law electron distribution.

For stars with X-ray-derived coronal temperatures, a thermal GS detection can characterize the coronal magnetic field or provide an upper limit for non-detection. Likewise, thermal GS measurements can estimate coronal temperature for stars with measured coronal magnetic field strengths and extent derived from, for example, Zeeman–Doppler imaging (Donati & Semel 1990). A few previous papers have reported the detection of thermal GS emission, but mainly in solar flares (Dulk, Melrose & White 1979; Kobayashi et al. 2006; Tan 2022). Drake et al. (1992) considered whether thermal GS can explain the spectral properties of low-level radio emission from RS CVn binary stars but reached no conclusion.

★ E-mail: wgolay@uiowa.edu (WWG); robert-mutel@uiowa.edu (RLM)

In this paper, we report on a search for thermal GS emission from five radio-loud stars that we selected based on previously published detection of high coronal temperature and strong magnetic fields. The stars comprised two classes: chromospherically active binaries (Algol, UX Arietis, HR 1099) and weak-lined T Tauri (WTT) pre-main-sequence stars (V410 Tau, HD 283572). The observations sample the SEDs in Stokes I and V over 15 – 45 GHz. We fit the observed SEDs with model SEDs generated by a two-component emission model consisting of a mildly relativistic population with a power-law electron energy distribution and a hot thermal component with a Maxwellian energy distribution. Each model had five free parameters that parametrized the fit to the observed SEDs. For convenience, we include the PYTHON code used to generate each figure, found on this Github repository and archived at doi:10.5281/zenodo.7783327.

## 2 GS RADIATION: SUMMARY OF PROPERTIES

### 2.1 Calculation of emergent model flux for a uniform plasma

The derivation of the volume emission ($\eta_\nu$) and linear absorption ($\kappa_\nu$) coefficients for GS radiation, whether by power-law or thermal electron distributions, is notoriously difficult, since it involves integrals over infinite sums of Bessel functions. However, several clever approximation methods have been developed that have made this calculation tractable (e.g. Trubnikov 1958; Dulk et al. 1979; Petrosian 1981; Leung, Gammie & Noble 2011). These methods have since developed into comprehensive packages for rapidly calculating SEDs from arbitrarily defined energetic particle distributions, enabling the evaluation of complex, strongly inhomogeneous models (e.g. Fleishman & Kuznetsov 2010; Kuznetsov & Fleishman 2021).

Robinson & Melrose (1984) derived analytic expressions for GS coefficients $\eta_\nu$ and $\kappa_\nu$ for both power-law and thermal electron distributions based on these approximations. We have used these expressions to calculate the expected SED and polarization for a homogeneous magneto-active plasma volume with electron energy populations consisting of (i) a hot isothermal plasma or (ii) a plasma with a power-law energetic electron population viz.,

$$n_p(E)dE = n_e \cdot \frac{\delta - 1}{E_0} \left[\frac{E}{E_0}\right]^{-\delta} dE, \quad (1)$$

where $n_e$ is power-law electron density and $E_0$ is the minimum cutoff energy, we assume $E_0 = 10$ keV. For each volume, we solve the equation of radiative transfer along the lines of sight intercepting the coronal plasma.

For a uniform homogeneous plasma, the radiative transfer equation can be easily integrated, so that emergent flux $S$ for a given emission mode may be written,

$$S_i = \mathscr{S}_i \cdot (1 - e^{-\tau_i}) \cdot \Omega, \quad (2)$$

where $i$ is the emission mode index (ordinary or extraordinary more for power-law or thermal electron distribution function), $\mathscr{S}$ is the source function, $\Omega$ is the source solid angle as seen by the observer, and $\tau_i = \kappa_i \cdot L$, is the optical depth, where $\kappa_i$ is the linear absorption coefficient for mode $i$ and $L$ is the source extent along the observer's line of sight. The source function is given by

$$\mathscr{S}_i = \frac{\eta_i}{\kappa_i}, \quad (3)$$

where $\eta_i$, $\kappa_i$ are the volume emission and linear absorption coefficients for mode $i$.

Robinson & Melrose (1984) derive expressions for these coefficients as a function of plasma parameters for the case of power-law (equation 52a, b) and thermal electron distributions (equations 42, 43), which we used in to calculate model fluxes. Note that the derivation of these expressions assumes only that the frequency of emergent radiation is much greater than both the source electron plasma and cyclotron frequencies, but *not* that the source is optically thin. Also, we do not include the contribution of bremsstrahlung (free–free) radiation since in all cases considered in this work, bremsstrahlung contributes less than 1 per cent of the total flux density.

### 2.2 Maxwellian electron distributions

The SED for radiation emitted by a non-relativistic thermal plasma in a magnetic field ('gyro-resonance radiation') is the sum of contributions from individual electrons emitting at the local gyro frequency and the first few harmonics. The radiation is largely circularly polarized at aspect angles outside the plane of rotation. However, as the plasma temperature exceeds $T_e \sim 10^7$ K, a significant fraction of the electron population becomes relativistic. This case significantly modifies both the SED and polarization. At large optical depth, the SED is a power law with spectral index $\alpha = +2$ and is steeply negative ($\alpha \sim -15$) at a small optical depth. The peak flux occurs near $\tau = 2.5$, where $\tau$ is approximately given by,

$$\tau_\nu \sim 1.2 \cdot \left[\frac{T_e}{10^8 \text{ K}}\right]^7 \cdot \left[\frac{B}{\text{kG}}\right]^9 \cdot \left[\frac{\nu}{10 \text{ GHz}}\right]^{-10} \cdot \left[\frac{n_e}{10^5 \text{ cm}^{-3}}\right] \cdot \left[\frac{L}{R_\odot}\right], \quad (4)$$

where $T_e$ is the plasma kinetic temperature, $B$ is the magnetic field strength, $\nu$ is the frequency, $n_e$ is the thermal electron density, and $L$ is the depth along the line of sight to the observer.

The frequency at SED maximum can be written approximately as

$$\nu_{\text{peak}} \sim 10 \text{ GHz} \cdot \left[\frac{B}{\text{kG}}\right] \cdot \left[\frac{T_e}{10^7 \text{ K}}\right]^{0.5} \cdot \sin^{0.6}\theta \quad (5)$$

(recast equations 32a, b Dulk 1985). Note that the peak frequency depends on the magnetic field and plasma temperature and is insensitive to density or path length.

The polarization is elliptical, but the axial ratios for ordinary and extraordinary modes are close to unity unless the propagation direction is nearly perpendicular to the magnetic field. This effect results in circularly polarized modes for propagation angles $\theta$ that satisfy (Robinson & Melrose 1984),

$$|\theta - \pi/2| \gg \frac{\nu_B}{2\nu}, \quad (6)$$

where $\nu$ is the emission frequency and $\nu_B$ is the electron gyro-frequency. The polarization fraction is relatively high, typically 40 – 90 per cent for a homogeneous plasma near the peak emission frequency, but decreases sharply with increasing optical depth and is nearly zero for $\tau \gg 1$.

Inspection of equations (4) and (5) indicate that we expect thermal GS radiation to be most prominent for very hot (>10 MK) coronal plasmas with strong ($\sim$kG) magnetic fields. The SEDs will have spectral peaks above 10 GHz and be circularly polarized. Sample model thermal GS spectra (Stokes I and V/I) are shown in Fig. 1 for uniform magnetic fields of 0.5 and 1.5 kG, and for plasma temperatures $10^{7.5}$ and $10^8$ K.

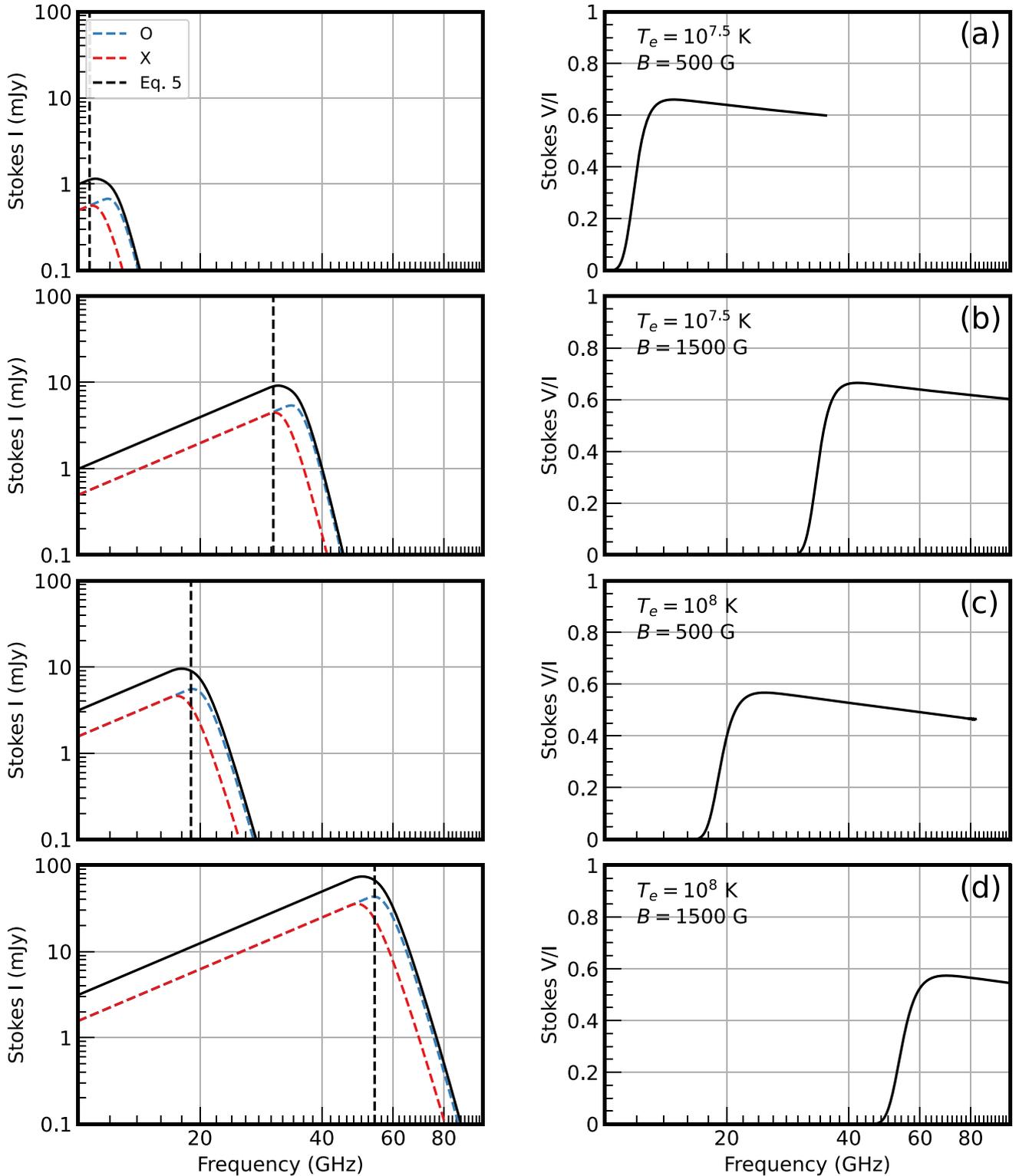

**Figure 1.** Stokes I spectral energy density distributions (left panels) and fractional circular polarization (right panels) for thermal GS emission from a homogeneous stellar corona modelled as a uniform cube with sides $1\,R_\odot$ at a distance 10 pc, with constant electron density $n_e = 10^{10}\,\mathrm{cm}^{-3}$, and uniform magnetic field oriented $80°$ inclination to the observer's line of sight. (a) $B = 500$ G, $T_e = 10^{7.5}$ K, (b) $B = 500$ G, $T_e = 10^8$ K, (c) $B = 1500$ G, $T_e = 10^{7.5}$ K, and (d) $B = 1500$ G, $T_e = 10^8$ K. The Stokes I plot lines are X-mode (dashed-red line), O-mode (dashed-blue line), and total thermal GS emission (solid-black line). The peak frequency as predicted by equation (5) is plotted (vertical-dashed black line). The cutoff in circular polarization at high frequencies is due to the numerical instability as the flux in the individual modes approaches zero.

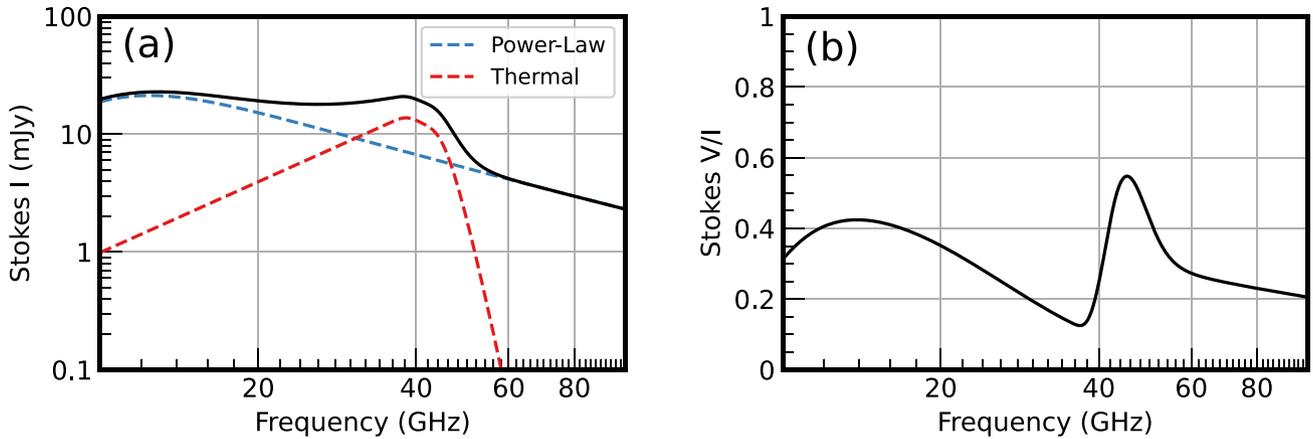

**Figure 2.** (a) Stokes I SEDs for the sum of two uniform plasma volumes with different emission processes: power-law GS emission (dotted-blue line) and thermal GS emission (dotted-red line). Both plasma models are uniform cubes with sides 1 $R_\odot$ at 10 pc distance. The power-law region has density $n_e = 10^7$ cm$^{-3}$, energy index $\delta = 3$, and a $B = 200$ G magnetic field. The thermal box has an electron density $n_e = 10^{10}$ cm$^{-3}$, a temperature $T_e = 10^{7.5}$ K, and a $B = 2$ kG magnetic field strength. The magnetic field is oriented at 45° and 60° to the observer's line of sight, respectively. (b) The corresponding fractional circular polarization (Stokes V/I) of the combined emission demonstrates the relatively narrow but substantial peak characteristic of thermal GS emission near the emission peak frequency.

### 2.3 Power-law electron distributions

The SED of radiation by highly relativistic power-law electrons in a magneto-plasma is well-known, consisting of a rising spectrum with $\alpha = +5/2$ at large optical depth, a falling spectrum with $\alpha = (1 - \delta)/2$ at small optical depth, and a peak near $\gamma^2 \nu_B$, where $\gamma$ is the Lorentz factor and $\nu_B$ is the gyro frequency. Relativistic beaming confines the radiation to small angles perpendicular to the magnetic field, resulting in linear polarization.

The SED and polarization characteristics are significantly different for mildly relativistic power-law electron distributions (frequencies roughly 10x – 100x the electron gyrofrequency). At large optical depths, the SED is also a power law but with a somewhat steeper slope ($\alpha = 2.5 + 0.085\delta$, Dulk 1985). At small optical depth, the slope $\alpha$ is also somewhat steeper than the fully relativistic case,

$$\alpha \sim 1.6 - \frac{(1-\delta)^{1.25}}{2}. \quad (7)$$

Power-law GS radiation is circularly polarized for propagation angles that satisfy equation (6). For a uniform magnetic field and $\tau \ll 1$, the fractional polarization varies from V/I ∼ 0.9 at low harmonics and steep power-law index to V/I < 0.1 for high harmonics and shallow index. These fractions pertain to plasmas with unidirectional magnetic fields. Of course more realistic magnetic geometries will result in lower fractions.

### 2.4 SED and polarization for combined thermal and power-law GS emission

In this paper, we will consider whether observed SEDs from stellar coronae result from GS emission from power-law electrons (as is often invoked), or whether the spectra are a composite of emission from different regions dominated by power-law and hot thermal electrons. Fig. 2 shows a representative composite spectrum consisting of the sum of SEDs from two equal spatially distinct volumes, a cube with dimension one solar radius on each side:

(i) a power-law GS emission region with parameters: $n_e = 10^7$ cm$^{-3}$, $\delta = 3$, $B = 200$ G; dashed-blue line) with a turnover near unity optical depth, and

(ii) a thermal GS emission from a denser region of hot thermal plasma and high magnetic field strength ($n_e = 10^{10}$ cm$^{-3}$, T = $10^{7.5}$ K, $B = 2$ kG; dashed-red line).

For this example, the thermal GS peak is somewhat near the power-law SED peak and has a similar peak flux density, so it may be challenging to distinguish between these emissions. However, a key difference is the degree of circular polarization, which sharply peaks near the thermal GS spectral peak. The key distinguishing feature of thermal GS emission is this sharp rise in fractional circular polarization near the spectral peak.

## 3 OBSERVATIONS AND DATA REDUCTION

We observed five stars with previous radio detections. We selected the target list based on two criteria: (i) The star had sufficient X-ray spectral information to be consistent with a hot component $T_e \sim 10^7$ K or higher, (ii) The star had evidence for ∼ kG magnetic fields, inferred from Zeeman–Doppler imaging (ZDI) or non-thermal radio emission.

We observed each star with the JVLA using 6 frequency bands spanning a total of 2 h with the following cadence: Ku band (14 – 16 GHz), 3 min; K band (22 – 24 GHz), 6 min; Ka lower band (29 – 30 GHz and 32 – 33 GHz), 6 min; Ka upper band (34 – 36 GHz), 9 min; Q lower band (40 – 42 GHz), 15 min; and Q upper band (44 – 46 GHz), 18 min. We chose the varying integration times to account for the decreasing sensitivity of the JVLA at higher frequencies. The observations spanned a total of five 2-h intervals (Table 1). We used the CASA software suite (The CASA Team et al. 2022) to edit, calibrate, and image each field. We determined Stokes I and V fluxes using 2D Gaussian fits and corresponding uncertainties. Table 2 summarizes the resulting flux densities and uncertainties.

## 4 ANALYSIS: MODEL FITTING

To determine the physical properties of the plasma responsible for the observed emission, we fit the observed SEDs and polarizations by calculating the emergent flux and polarization for a uniform plasma

**Table 1.** Star physical properties, observing log, and priors.

| Star | Sp. type | Class | Distance[a] [pc] | Epoch | UT range | Size [$R_\odot$] | Ref.[b] | Prior values $L_X$ [$10^{31}$ erg s$^{-1}$] | Ref.[c] | B [kG] | Ref.[d] |
|---|---|---|---|---|---|---|---|---|---|---|---|
| HR 1099 | K1IV+G5V | RS CVn | 29.6 | 02 Oct 2013 | 02:30 – 12:30 | 7.2 | 1 | 1.3 | 4 | 1.0 | 6 |
| UX Arietis | K0IV+G5V | RS CVn | 50.5 | 24 Dec 2013 | 02:30 – 12:30 | 6.5 | 2 | 1.7 | 4 | – | – |
| Algol | K0IV+B8V | Algol | 27.6 | 29 Sep 2013 | 10:07 – 12:08 | (7.2)[e] | 3 | 1.3 | 4 | – | – |
| V410 Tau | K2 | WTTS | 128.7 | 08 Jan 2014 | 06:07 – 11:36 | – | – | 0.4 | 5 | 1.1 | 7 |
| HD 283 572 | G5IV | WTTS | 125.5 | 07 Jan 2014 | 05:33 – 07:33 | – | – | 1.2 | 5 | – | – |

*Notes.* [a] Distances are from the SIMBAD data base (Wenger et al. 2000).
[b] Size references: (1) Mutel et al. (in preparation), (2) Peterson et al. (2011), (3) Peterson et al. (2010)
[c] X-ray luminosity references: (4) Ness et al. (2002), (5) Telleschi et al. (2007)
[d] Magnetic field references: (6) Donati (1999), (7) Finociety et al. (2021)
[e] The prior for each lobe of the Algol model was half this value. See Section 5.1.3 for model details.

with prescribed physical parameters, adjusting each parameter to fit the observed fluxes. We adopted a simple geometric cube box model consisting of a homogeneous population of electrons described by either a power-law or a thermal energy distribution. The emitting plasma was parametrized by a characteristic size $L$, number density of electrons $n_e$, magnetic field strength $B$, and magnetic field angle $\phi$. A power-law index $\delta$ and the electron temperature $T_e$ define the power-law and thermal regions, respectively. We reconstructed model Stokes I and V fluxes by summing and differencing the model's O- and X-mode emission, respectively. As mentioned earlier, this assumes that the propagation angle is not too close to the magnetic field perpendicular direction (equation 6). Given these simplifying assumptions, the derived physical parameters should be interpreted as volume averages, whereas actual values could (and almost certainly do) vary within the emitting plasma.

In order to test whether a thermal GS component was present in the observed spectra, we fit two types of models: a pure power-law model and a hybrid model comprising the sum of a power-law region and a separate thermal region. For each star, we fit both power-law and hybrid models, and then calculated the Bayesian Information Criterion (BIC) to evaluate whether the addition of a second thermal component was statistically justified.

### 4.1 Bayesian inference

Since the number of independent data points for each star (10 – 12) is not much larger than the number of free parameters in the model (5 for power-law or 10 for hybrid models), the fitted parameters may be degenerate, which we would like to fully characterize. In addition, we would like to incorporate constraints on parameter values from previous observations, such as radio sizes from very long baseline interferometry (VLBI), X-ray luminosities, and magnetic field measurements from ZDI.

Bayesian inference is a natural choice here. First, although we cannot avoid degeneracies, we can better characterize them by numerically sampling the posterior distribution to determine the joint probability density of each parameter pair. Secondly, priors can naturally include information about parameter constraints from previously published studies. Thirdly, the BIC can be used to provide a metric for comparing models with different numbers of parameters, a critical test to evaluate whether the addition of a thermal component is statistically justified.

We implement a constraint on index of refraction $n$ on all the stars. Any plausible model must be free-space accessible ($n$ is real) across all the observed frequencies. To avoid using a prior since this not an independently measured quantity but rather a consequence of the model, we integrate this constraint directly into the model. We take the product of the emergent flux from the model with a Heaviside function evaluated on $n$ at the observed frequencies. In the case of imaginary $n$, the emission at that frequency is zero, and the likelihood function will evaluate that model as a poor fit. In preliminary models of V410 Tau and HD 283572, $n$ is imaginary up to the lowest observed frequency at 15 GHz. To counteract this non-physical edge case, we require that $n$ remain real as low as 3 GHz so as not to obscure flux detected from these two stars in the Very Large Array Sky Survey from the same emitting plasma (Lacy et al. 2020; Gordon et al. 2021).

Stars with previous measurements of the plasma parameters via an independent method have a prior implemented unique to the star. The values we select are detailed in Table 1. In all cases, the prior probability distribution function (PDF) is a Gaussian centred on the value with a 20 per cent width. The parameters include

(1) Characteristic size: for the three close active binaries, VLBI measurements were available to constrain the characteristic size $L$ of the power-law region. The Algol dual-lobe total extent was reported as 7.2 $R_\odot$, so we enforced a Gaussian prior of 20 per cent width centred at half that value for each power-law region in the Algol model (see Section 5.1.3 for model details).

(2) Thermal plasma X-ray emission: we approximate the X-ray luminosity $L_X$ of a hot ($T_e \gg 1$ keV) hydrogen plasma by determining the integrated bremsstrahlung (free–free continuum emission) for a uniform cube with side $L_{\rm box}$ (e.g. Karzas & Latter 1961, equation 26, recast)

$$P_{\rm Br} = 1.5 \times 10^{29} \cdot \left[\frac{n_e}{10^{10}\,{\rm cm}^{-3}}\right]^2 \cdot \left[\frac{T_e}{10^7\,{\rm K}}\right]^{0.5} \cdot \left[\frac{L_{\rm box}}{R_\odot}\right]^3 {\rm erg\ sec}^{-1}, \quad (8)$$

where $n_e$ and $T_e$ are the number density and electron temperature of the thermal population. We compare the expected X-ray luminosity of the thermal plasma model with existing estimations.

(3) Magnetic field strength: a 6-yr ZDI campaign of HR 1099 demonstrated local peaks of up to $\sim 1$ kG. ZDI of V410 Tau measured a magnetic topology with local radial components up to 1.1 kG. We use these values as the prior for the thermal population since any substantial emission is likely to come from the regions with the strongest fields (Fig. 1).

### 4.2 Fitting procedure

To determine an optimized model for each target, we followed a six step process:

**Table 2.** Observed Stokes flux densities.

| Star | Stokes | 15 GHz | 23 GHz | 31 GHz | 35 GHz | 41 GHz | 45 GHz |
|---|---|---|---|---|---|---|---|
| HR 1099 | I | 20.90 ± 0.09 | 14.66 ± 0.10 | 9.99 ± 0.09 | 8.84 ± 0.06 | 7.26 ± 0.08 | 6.41 ± 0.07 |
|  | V | 6.55 ± 0.04 | 5.05 ± 0.04 | 3.04 ± 0.03 | 2.52 ± 0.03 | 2.00 ± 0.05 | 1.63 ± 0.05 |
| UX Arietis | I | 78.82 ± 0.29 | 57.44 ± 0.17 | 43.62 ± 0.17 | 38.79 ± 0.12 | 30.72 ± 0.22 | 31.13 ± 0.17 |
|  | V | −11.98 ± 0.05 | −8.33 ± 0.03 | −4.81 ± 0.04 | −3.84 ± 0.03 | ‡ | −2.32 ± 0.05 |
| Algol | I | 41.83 ± 0.18 | 32.82 ± 0.16 | 25.98 ± 0.19 | 25.63 ± 0.13 | 23.97 ± 0.22 | ‡ |
|  | V | −0.32 ± 0.07 | −0.48 ± 0.03 | −0.49 ± 0.04 | −0.75 ± 0.06 | −0.500 ± 0.07 | ‡ |
| V410 Tau | I | 1.27 ± 0.03 | 1.28 ± 0.02 | 1.11 ± 0.03 | 1.08 ± 0.02 | 0.80 ± 0.03 | 0.76 ± 0.04 |
|  | V | −0.11 ± 0.02 | −0.13 ± 0.01 | −0.14 ± 0.02 | −0.17 ± 0.02 | −0.19 ± 0.02 | −0.23 ± 0.03 |
| HD 283572 | I | 2.29 ± 0.02 | 1.89 ± 0.02 | 1.44 ± 0.03 | 1.13 ± 0.02 | 0.97 ± 0.03 | 0.82 ± 0.04 |
|  | V | 0.27 ± 0.04 | 0.24 ± 0.04 | 0.27 ± 0.03 | 0.24 ± 0.03 | 0.28 ± 0.05 | 0.21 ± 0.05 |

*Note.* All fluxes are in mJy. ‡ indicates missing observation.

(1) Initialize a model by selecting either a pure power-law electron population or the sum of a power-law and a non-cospatial hot thermal region with parameters informed by literature-reported values (see Table 1).

(2) Search for best-fitting parameter values using a downhill simplex algorithm (Nelder–Mead) to minimize a weighted $\chi^2$ objective function. The PYTHON package lmfit (Newville et al. 2014) was used for this step.

(3) Initialize the relevant priors (characteristic size, X-ray temperature, magnetic field measurements) for the selected model and target.

(4) Invoke Bayesian inference via Markov Chain Monte Carlo (MCMC) sampling using PYTHON package EMCEE (Foreman-Mackey et al. 2013) to determine best-fitting parameter values and associated uncertainty PDFs.

(5) Test for walker convergence (see Appendix B for details) by:

(i) Determining the 'goodness-of-chain' by investigating the walker acceptance fraction and the total number of steps compared to the MCMC autocorrelation time as a function of step-index.

(ii) Adjusting the number of steps or the stretch step parameter and re-running the chain as needed.

(6) Compare BICs for pure power-law versus hybrid models with separate power-law and thermal components to determine whether the addition of a thermal component is statistically justified.

We modelled the observed SEDs using both power-law and the sum of power-law plus thermal GS emission models in separate, non-cospatial regions. However, for Algol we modelled two oppositely polarized power-law emission regions based on a previous VLBI map (Mutel et al. 1998 showed spatially distinct lobes of opposite helicity). The fitting procedure began by applying a downhill simplex algorithm with a $\chi^2$ objective function. The resulting best-fit model was used as a starting point for initializing the Bayesian MCMC analysis. Priors for the VLBI sizes, X-ray luminosity, and magnetic field were set based on literature-reported values as listed in Table 1, and any others were set to default prior settings (see Appendix A for the default priors and a discussion of the implementation of PDFs). To encode observational uncertainties into the model-fitting procedure, the objective likelihood function was the residual sum of squares between the model and observed fluxes weighted by their uncertainties ($\chi^2$).

For each MCMC calculation, we initialized 100 walkers in a small (1 per cent of parameter values) Gaussian hypervolume around the minimized solution. Our step proposal density function was the 'stretch-step' algorithm (Goodman & Weare 2010), a variable step-length algorithm designed to quickly fill posterior space by scaling step sizes with the distribution of walkers. For all runs, a period of tracked 'burn-in' steps gave the walkers time to undergo this filling process.

To determine if the walkers had converged, we evaluate the 'goodness-of-chain' of the MCMC run based on the walker acceptance fraction and the autocorrelation time compared to the total step number. For chains with acceptance fractions below 0.1 or above 0.5, the average step length was shortened or lengthened, respectively, with a target acceptance fraction of 0.23 (Gelman, Gilks & Roberts 1997). We define convergence of the chain as the total number of steps exceeding 50 times the autocorrelation time of the chain as estimated by EMCEE. We increased the length until the chain met this criterion. See Appendix B for an extended discussion of evaluating the goodness-of-chain.

### 4.3 Testing the thermal GS hypothesis: BIC

Since a primary focus of this work is to determine whether thermal GS radiation is detectable in the observed SEDs, we evaluated whether the addition of a thermal GS component to the emission model was statistically justified. To do this, we use the BIC, defined as

$$\mathrm{BIC} = k\ln(n) - 2\ln(\hat{L}), \qquad (9)$$

where $k$ is the number of parameters estimated by the model, $n$ is the number of data points in the observations, and $\hat{L}$ is the maximized value of the likelihood for that particular model, that is, $\hat{L} = p(x|\hat{\theta}, M)$, where $\hat{\theta}$ are the parameters that maximize the likelihood and $x$ is the observed data (Schwarz 1978). The BIC can be used to select the model that fits the data with a minimum number of free parameters by penalizing model complexity, that is, the number of parameters in the model (Liddle 2007).

For each star, we evaluated the BIC for a pure power-law model, and for the hybrid model. Although the actual BIC value depends on the particular values of $n$, $k$, and $\hat{L}$ for that model, the key metric is the difference in BIC values between two models,

$$\Delta \mathrm{BIC} = \mathrm{BIC}_{\mathrm{pwr+th}} - \mathrm{BIC}_{\mathrm{pwr}}, \qquad (10)$$

where $\Delta \mathrm{BIC} < 0$ favours the inclusion of the additional free parameters. $\Delta \mathrm{BIC}$s between -2 and -6 are considered 'positive' evidence for the more complex model, whereas a difference between -6 and -10 constitutes 'strong' evidence, and $\Delta \mathrm{BIC}$s beyond this are 'very strong' evidence (Bauldry 2015).

We find that the three close binaries (Algol, HR 1099, UX Arietis) have large positive $\Delta \mathrm{BIC}$s. indicating that a simple power-law model is adequate, whereas the two pre-main-sequence stars (V410 Tau,

**Table 3.** Best-fitting model parameters.

| Star | Power-law component | | | | | Thermal component | | | | | Selection metric[a] |
|---|---|---|---|---|---|---|---|---|---|---|---|
| | $L$ [$R_\odot$] | $\delta$ | $\log n_e$ [cm$^{-3}$] | $B$ [G] | $\phi$ [deg] | $L$ [$R_\odot$] | $\log T$ [MK] | $\log n_e$ [cm$^{-3}$] | $B$ [G] | $\phi$ [deg] | $\Delta$BIC |
| HR 1099 | 2.6 | 3.1 | 6.6 | 240 | 122° | – | – | – | – | – | 8.6 |
| UX Arietis | 7.8 | 2.7 | 5.7 | 180 | 85° | – | – | – | – | – | 11.8 |
| Algol | 3.4 | 2.1 | 4.7 | 170 | (54°, 146°)[b] | – | – | – | – | – | – |
| V410 Tau | 1.1 | 3.0 | 8.5 | 80 | 65° | 0.6 | 8.3 | 10.7 | 1110 | 31° | −15.4 |
| HD 283 572 | 2.6 | 3.6 | 8.6 | 110 | 110° | 0.6 | 8.6 | 10.9 | 740 | 157° | −21.2 |

*Note.* Parameter uncertainties are not listed in this table. They are best characterized by the joint posterior model parameter probability distributions shown in Fig. 4 for HR 1099 and UX Arietis, Fig. 5 for Algol, Fig. 7 for V410 Tau, and Fig. 8 for HD 283572. The listed model parameters are the median values of the MCMC sampling. [a] The selection metric is defined as difference between the BICs of the power-law GS model and the power-law plus thermal GS model. [b] The Algol model requires two angles that parameterize the two oppositely circularly polarized emission lobes (Section 5.1.3).

HD 283572) have large negative $\Delta$BICs, indicating that the more complex hybrid model is favoured. The $\Delta$BICs for all stars are listed in Table 3.

## 5 RESULTS

Here, we summarize the results of the fitting procedure. For each star, we overlay a sample of the best-fitting models onto the measured SED in Stokes I and V/I (Fig. 3, 5a, and 6). We include the joint posterior model parameter probability distributions (corner plots in Fig. 4, 5b, 7, and 8) to represent the walker distribution in parameter hyper-space. This distribution is an estimate of the posterior space value, so we may use these 2D 'slices' of the walker density to expose cross-parameter correlations and degeneracies. We compare best-fit model plasma parameters (magnetic field, electron density, source size, etc.) with estimates from previous studies using other techniques. However, for all five stars the joint posterior probability distributions are strongly pairwise degenerate in at least some parameters. Hence, the parameter uncertainties, as shown in the parameter distribution functions about the median value, are not normally distributed, but are significantly skewed.

The derived best-fit values for physical parameters such as magnetic field strength or electron density are weighted averages over the radio emitting volume. As such, it may be problematic to compare these values to those using techniques that use different weighting and/or sampled volumes in the stellar environment. For example, radio-derived magnetic field strengths are volume averages sampled in the radio emitting plasma (presumably the extended stellar corona), whereas ZDI probes the stellar photosphere. Likewise, radio-derived source sizes are effectively the transverse extent of emitting plasma strongly weighted by magnetic field strength, electron density distribution or (for thermal GS) temperature, whereas X-ray-derived volumes depend only on temperature and density via the volume emission measure, which has no magnetic field dependence. Nevertheless, it is useful to compare radio-derived physical parameters to values obtained using other techniques as long as the different weighting dependencies are accounted for.

Our main focus is to establish whether there is evidence for thermal GS emission, that is, whether a pure power-law model alone or a hybrid model with a thermal GS component is required to fit the observed SEDs. To this end, we calculate the $\Delta$BIC for both models. As described earlier, this difference is a quantitative measure of the viability of the more complex hybrid model for each star.

Model fits for the three close active binaries have large positive values of the $\Delta$BIC metric (Table 3). This is a strong indicator that the addition of additional parameters beyond the power-law model is not justified, that is, that adding additional parameters would overfit the data set. In contrast, the best-fitting solutions for the hybrid (power-law versus power-law plus thermal) models of both pre-main-sequence stars have very negative BIC values (Table 3). This strongly favours the addition of a thermal component to explain the observed SEDs, particularly the rising fractional polarization at higher frequencies. In both cases, the fitted thermal emission regions have smaller volumes and are much denser than the power-law emission regions. Additionally, the best-fitting magnetic field strengths in the thermal regions are much larger (7x, 14x) than the power-law regions. These differences indicate that the two plasma populations cannot be co-spatial.

### 5.1 Active close binaries

#### 5.1.1 HR 1099

HR 1099 is a well-known active close binary with extensive studies at radio, UV and X-ray wavelengths. It was one of the earliest targets of ZDI investigations (Donati et al. 1990). More extensive ZDI monitoring (Donati 1999) found photospheric values of a few hundred Gauss extending over a large fraction of the photosphere, but with smaller regions exceeding 1 kG. This is consistent with our best-fit radio coronal value (240 G) but a detailed comparison would require knowledge of the magnetic topology.

A pure power-law electron distribution model fits the observed SEDs quite well (Fig. 3), obviating the need for a thermal GS component. This is supported by the large positive $\Delta$BIC = 8.6. The model predicts a flip to a negative Stokes V below $\sim$8 GHz, which agrees with the recent ASKAP detection at 888 MHz (Pritchard et al. 2021). The model angular size $L = 2.5\,R_\odot$ also agrees with the upper limit $L < 5\,R_\odot$ from VLBI at 22 GHz (Abbuhl & Mutel, in preparation) during a non-flaring state.

#### 5.1.2 UX Arietis

UX Arietis is another well-studied close active binary with numerous studies at optical, radio, and X-ray wavelengths. Donati, Semel & Rees (1992) reported early multi-epoch ZDI observations of UX Arietis. They detected circularly spectral profiles from which they inferred the presence of photospheric magnetic fields of order 100 G. Torricelli-Ciamponi et al. (1998) fitted radio spectra at several epochs using a power-law GS model. They found average magnetic field strengths in the range $300 \le B \le 1000$ G, power-law spectral indices $1.7 \le \delta \le 2.0$ and number densities $3.4 \le \log(n_e) \le 5.7$ cm$^{-3}$. Peterson et al. (2011) used a global VLBI network to map the radio structure at 15 GHz at several epochs. They found the radio structure partially resolved, corresponding to a source size $4.4\,R_\odot \times 8.8\,R_\odot$.

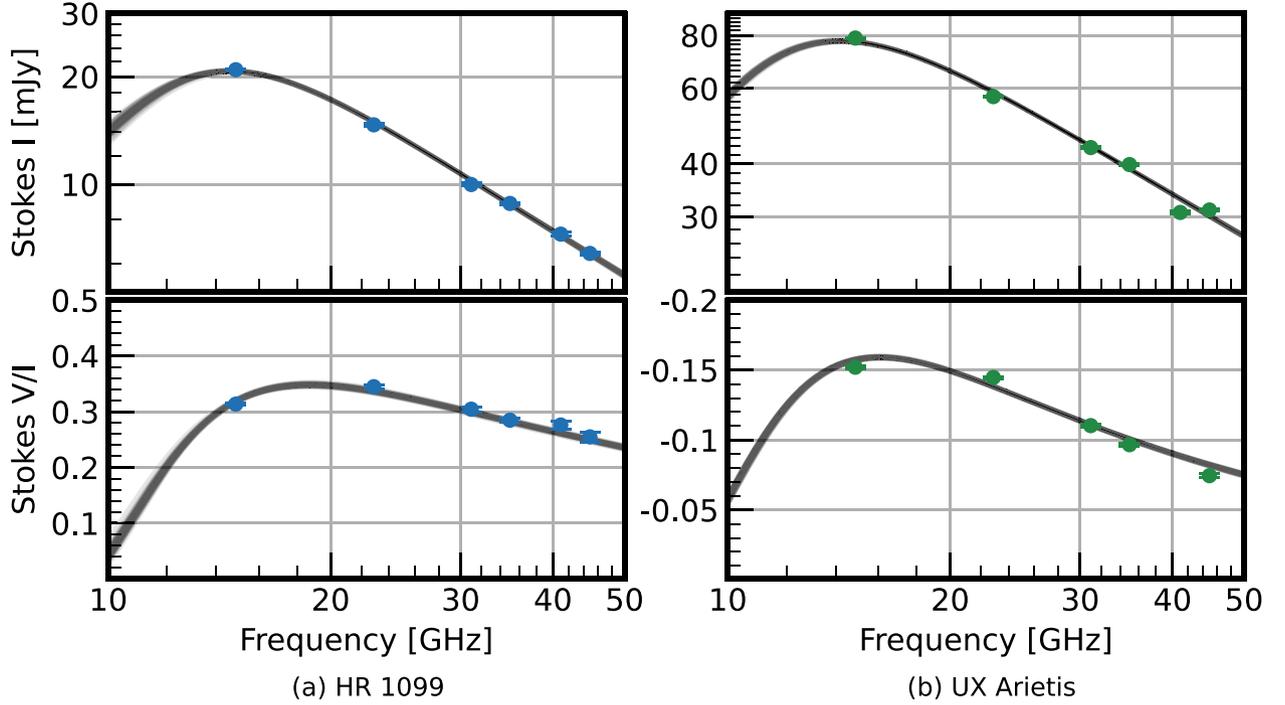

**Figure 3.** Observed SEDs for the close active binaries (a) HR 1099 and (b) UX Arietis. The best-fitting model SEDs are generated from 100 randomly selected walker positions in the final 10 per cent of the MCMC run. Some uncertainty bars are too small to resolve visually, but are listed in Table 2. Best-fitting parameters are listed in Table 3.

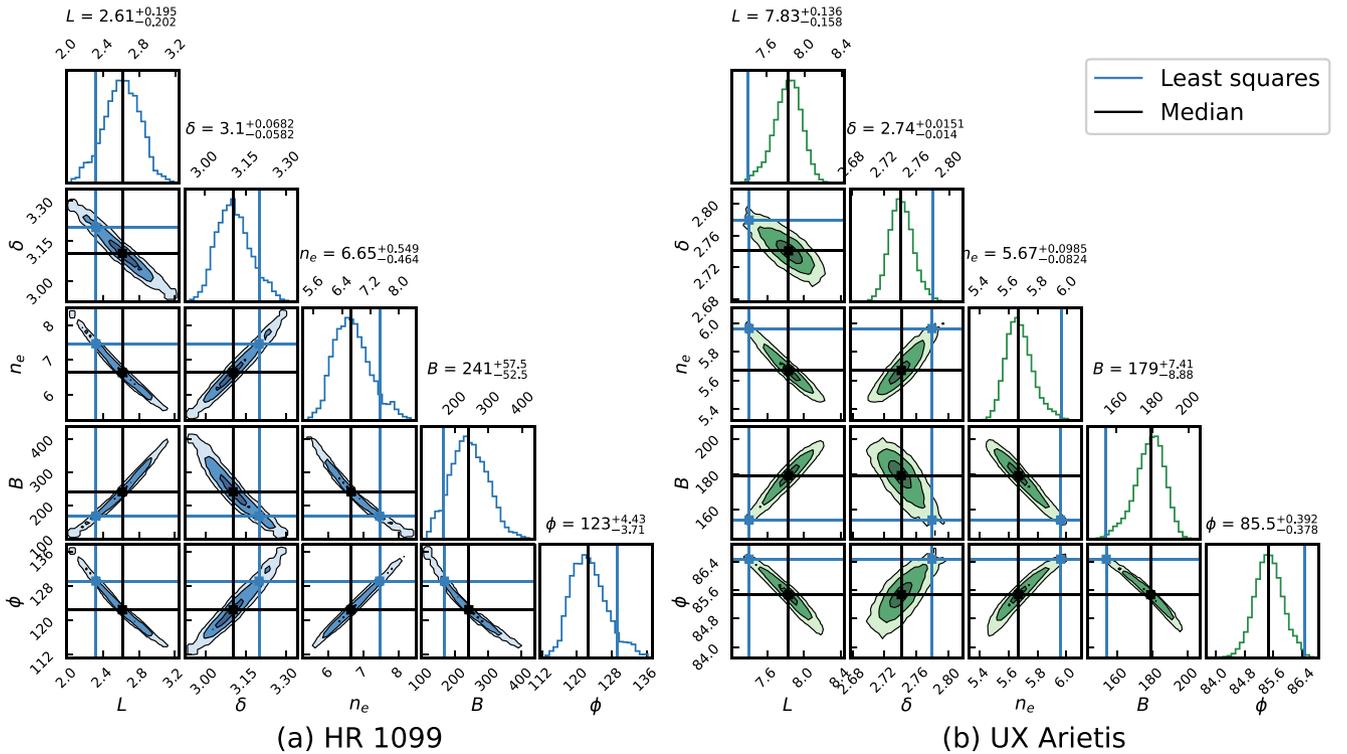

**Figure 4.** Joint posterior distributions for power-law electron distribution GS emission model fitted to SED of (a) HR 1099 and (b) UX Arietis. Note that most parameters are highly correlated, so unique values are degenerate. Contour levels are shown at 39, 87, and 99 per cent. The limits in the title of each column are the 16 and 84 percentiles. The legend highlights the best-fitting parameters reported by `lmfit` compared to the median value of the walker positions.

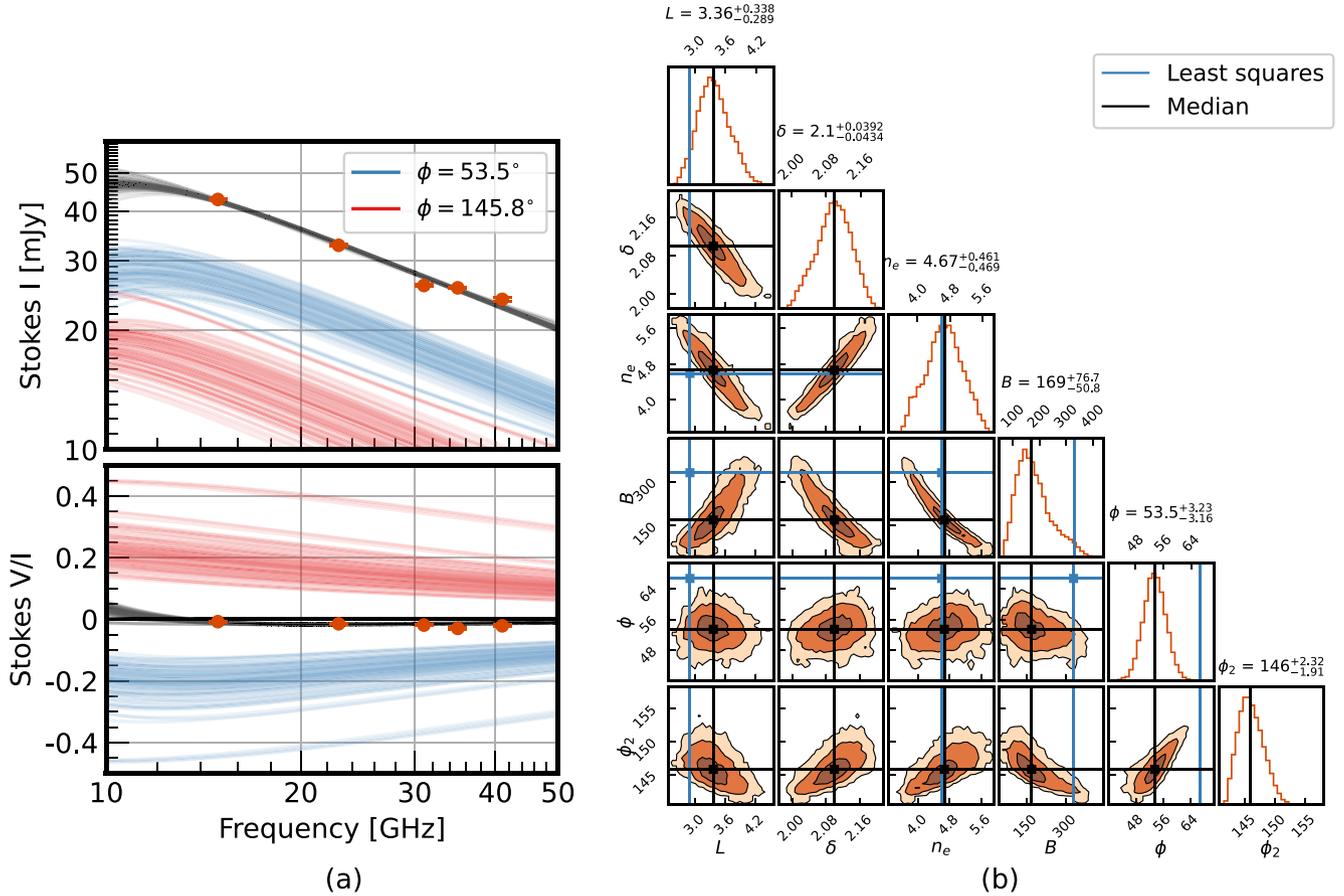

**Figure 5.** (a) Same as Fig. 3 and (b) Fig. 4 for Algol (see Section 5.1.3 for model details). The best-fitting models are plotted in black as the sum of the emission from each individual component specified by their angle, plotted in blue and red. Note that the low polarization is due to the opposite helicity of the two power-law regions. Some uncertainty bars are too small to resolve visually, but are listed in Table 2. Best-fitting parameters are listed in Table 3.

These independently derived physical parameters are all consistent with our best-fit values for UX Arietis given in Table 3.

The fit to the SED in Fig. 3 and the resulting $\Delta$BIC strongly support a pure power-law model with no evidence for a thermal GS component. As with HR 1099, the parameters are highly pairwise-degenerate, but are tightly constrained, with typical uncertainty ranges (39 – 87 per cent) of order 10 per cent of the best-fit values.

### 5.1.3 Algol

Algol, the prototype of the eponymous Algol class of close interacting binaries, exhibits activity at all wavelengths from radio to X-ray. Multiyear synoptic monitoring of the radio flux at cm wavelengths (Mutel et al. 1998; Retter, Richards & Wu 2005) demonstrates a basal level of order 10 mJy, with occasional outbursts as high as 1 Jansky. High resolution VLBI maps of the radio emission (Mutel et al. 1998) show that the emission arises from two well-separated lobes of opposite helicity. The lobes are approximately one subgiant diameter in height, with a base is straddling the subgiant and apex oriented towards the B8 star (Peterson et al. 2010). We modelled this morphology with two separate emission regions. Since the net circular polarization is nearly zero, the initial magnetic field angles were set in opposite directions.

For power-law model fitting, each emission region has five free parameters (size, power-law index, total electron density, magnetic field strength, and direction). Since in the case of Algol, we only obtained useful data at five frequencies (Table 2), the total number of degrees of freedom (5 + 5 = 10) would equal the number of data points, if all parameters will allowed to vary. Hence, in order to avoid zero degrees of freedom in the fitting procedure, we constrained the magnetic field strength, power-law spectral index, electron density, and size to be the same in each lobe, reducing the total number of free parameters to six (4 + 2 = 6). This is of course an ad hoc assumption, but is plausible given the observed similarities in size and flux density of the two lobes.

The two-component power-law model, shown in Fig. 5, provides a good fit to the observed SED, without the addition of a thermal component. This is supported by the highly positive $\Delta$BIC = 11.8. However, as noted earlier, the individual plasma parameters are somewhat degenerate, but the oppositely directed magnetic field orientation in the respective lobes is confirmed.

The model plasma parameters (Table 3) can be compared with previous estimates based in radio spectra: The model size of each lobe (3.4 $R_\odot$) agrees well with the VLBI-determined size, approximately half the total size of the two oppositely polarized lobes (Mutel et al. 1998). The model magnetic field (170 G), electron density (4.7 dex), and power-law slope (2.1) are somewhat different than those based on lower frequency synoptic Stokes I, V data Mutel et al. (1998) for the 'basal' component, but are comparable to the 'flare' component values ($B = 250$ G, $n_e = 3-5$ dex, $\delta = 3$). Since the present model is based on much denser and wider frequency sampling and has a more comprehensive uncertainty specification, it is probably a better estimator of the physical parameters of the emitting plasma than previously published values.

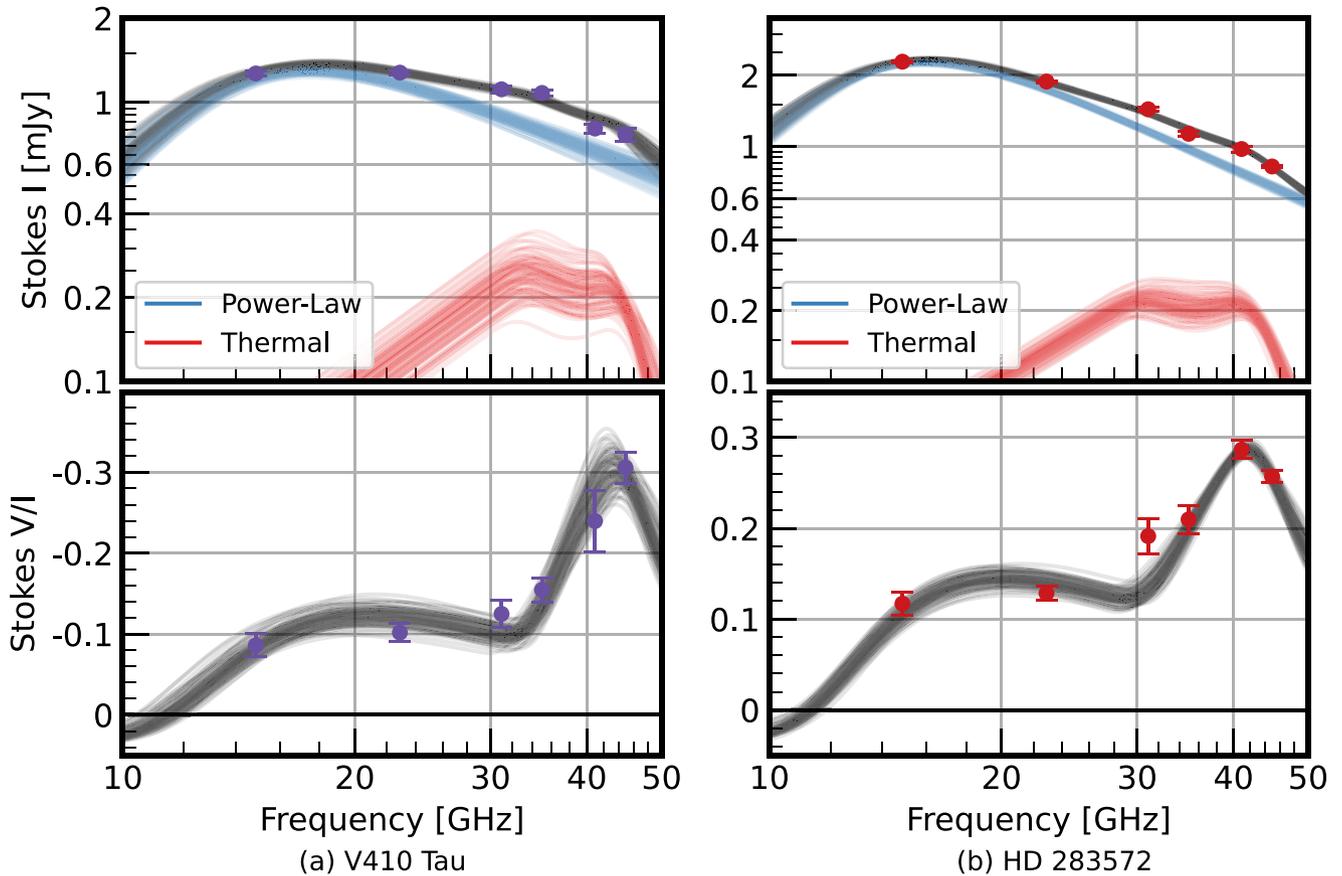

**Figure 6.** Same as Fig. 3 for the pre-main-sequence stars (a) V410 Tau and (b) HD 283 572 between 15 GHz and 45 GHz with best-fitting models consisting of separate power-law and hot thermal GS emission models. Some uncertainty bars are too small to resolve visually, but are listed in Table 2. Best-fitting plasma parameters are listed in Table 3.

### 5.2 WTT stars

*5.2.1 V410 Tau*

V410 Tau is a WTT in the nearby Taurus star formation region. It was one of the first stars whose magnetic fields were mapped via ZDI (Donati et al. 1997). A recent multiyear spectro-polarimetric imaging program (Finociety et al. 2021) revealed a complex, time-variable magnetic field topology. The poloidal component is non-axisymmetric and weakly dipolar while the toroidal component is mostly axisymmetric. The dipole component has a polar strength of 390 G with its axis tilted 15° to the rotation axis. The equally strong toroidal component remains to be explained in the absence of an observable disc. Telleschi et al. (2007) fitted the X-ray spectrum with a three-temperature composite model. They reported an average quiescent coronal temperature of 15 MK and the hottest component 24 MK.

The very large negative $\Delta$BIC $= -15.4$ (Table 3) is strong evidence that the addition of a thermal component is needed to fit the observed SED. The best-fit emission model comprises two distinct regions: (a) A $1.1\,R_\odot$ volume with a power-law electron distribution (3.0), $n_e \sim 10^{8.5}$ cm$^{-3}$, and mean magnetic field 80 G, and (b) a somewhat smaller ($0.6\,R_\odot$) volume with very hot ($10^{8.3}$ K) denser ($n_e \sim 10^{10.7}$ cm$^{-3}$) plasma and much stronger magnetic field $\sim 1.1$ kG.

This magnetic field strength in the thermal region is almost twice that of the largest values reported by Finociety et al. (2021) but is comparable to the maximum value (2 kG) found by Carroll et al. (2012). The X-ray luminosity as calculated using equation (8) and the best-fitting parameters gives $L_X = 3.4 \times 10^{30}$ erg sec$^{-1}$ and EM $= 8.6 \times 10^{53}$ cm$^{-3}$, in good agreement with Telleschi et al. (2007).

*5.2.2 HD 283572*

HD 283 572 is one of the brightest X-ray sources in the Taurus star-formation region (Favata, Micela & Sciortino 1998). X-ray measurements of quiescent emission indicate coronal temperatures of 19 − 26 MK (Franciosini et al. 2007), while fits to flare SEDs have an peak temperature ∼48 MK and magnetic fields 300 − 500 G assuming the hot plasma is confined in magnetic loops (Favata, Micela & Reale 2001). Phillips, Lonsdale & Feigelson (1991), using a VLBI network at 5 GHz, detected an unresolved ($\theta < 0.8$ mas) radio source, or a size $L < 0.12$ AU = 8.1 stellar radii, assuming a stellar radius 3.1 R$_\odot$ (Walter et al. 1987). Donati et al. (1997) using an early ZDI instrument, reported a 'marginal' detection of magnetic field but did not specify a field strength or topology. As with V410 Tau, no accretion disc has been detected (Sullivan & Kraus 2022).

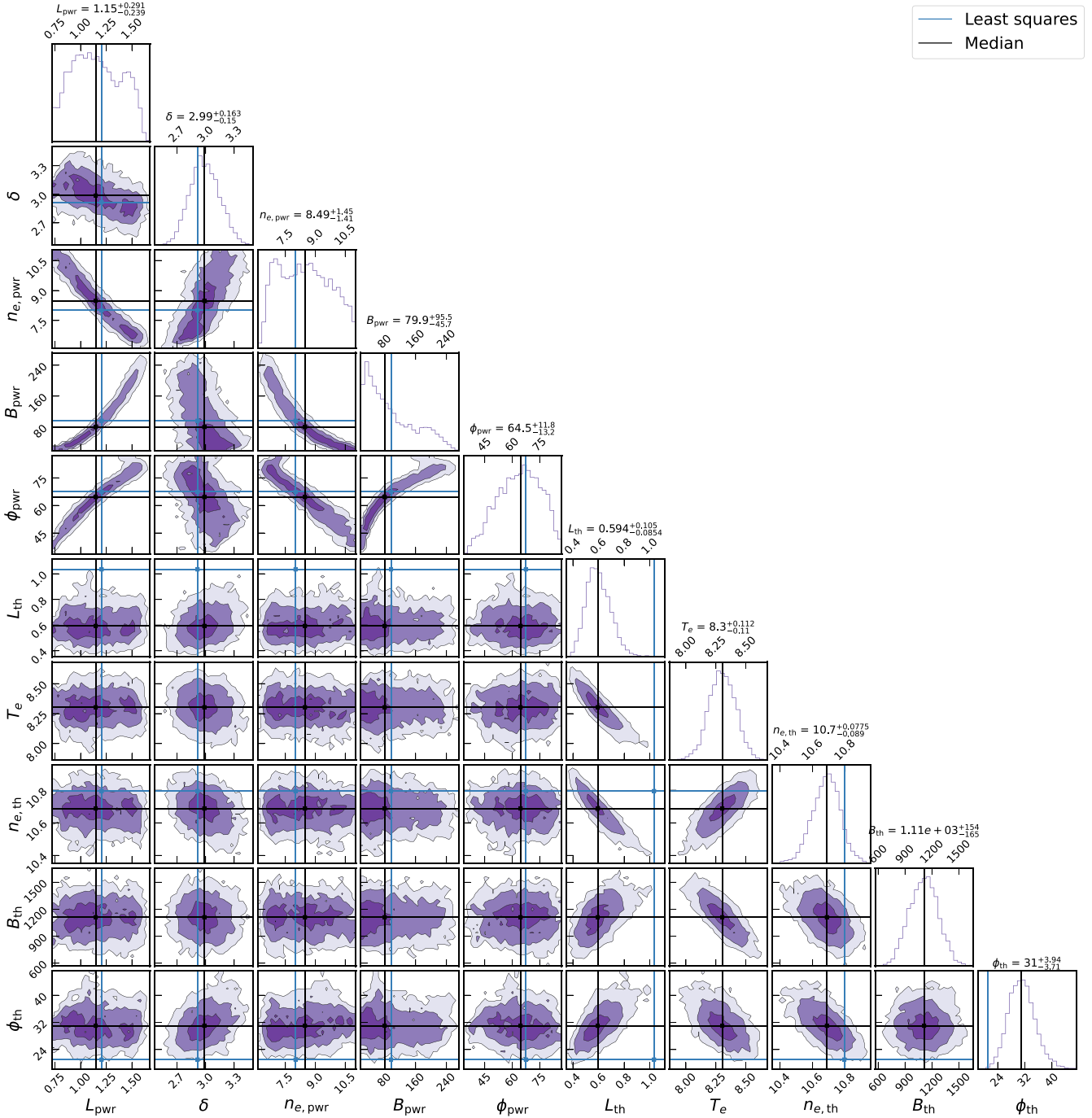

**Figure 7.** Same as Fig. 4 for V410 Tau. Correlations across the separate regions are much weaker compared to the degeneracies within the parameters that define each region. Note the general disagreement of the least squares result with nearly all of the thermal population plasma parameters.

The best-fit posterior model parameters, along pairwise-correlations between parameters, are shown in Table 3 and Fig. 8. The uncertainties are somewhat larger for the power-law model parameters compared to the thermal population. The X-ray luminosity prior may have constrained the thermal GS parameters somewhat more stringently than the power-law parameters. The model selection metric, $\Delta$BIC $= -21.2$, strongly favours a hybrid model with both power-law and thermal emission regions. As with previous models, all parameters are degenerate to some degree.

The power-law emission region size (2.6 R$_\odot$ = 0.8 stellar radii) is much larger than the thermal region (0.6 R$_\odot$ = 0.2 stellar radii) but much smaller than the upper limit determined from VLBI. The model has a mean magnetic field $B = 110$ G. This is significantly smaller than the inferred field in flaring loops, but is consistent with pressure confinement of the quiescent loops. The model volume emission measure (calculated from the mean density and volume) is in excellent agreement with the observed X-ray luminosity $L_X$ = $1.15 \times 10^{31}$ erg sec$^{-1}$ (Telleschi et al. 2007). Unfortunately, no direct ZDI measurements of the magnetic field strength or topology are available.

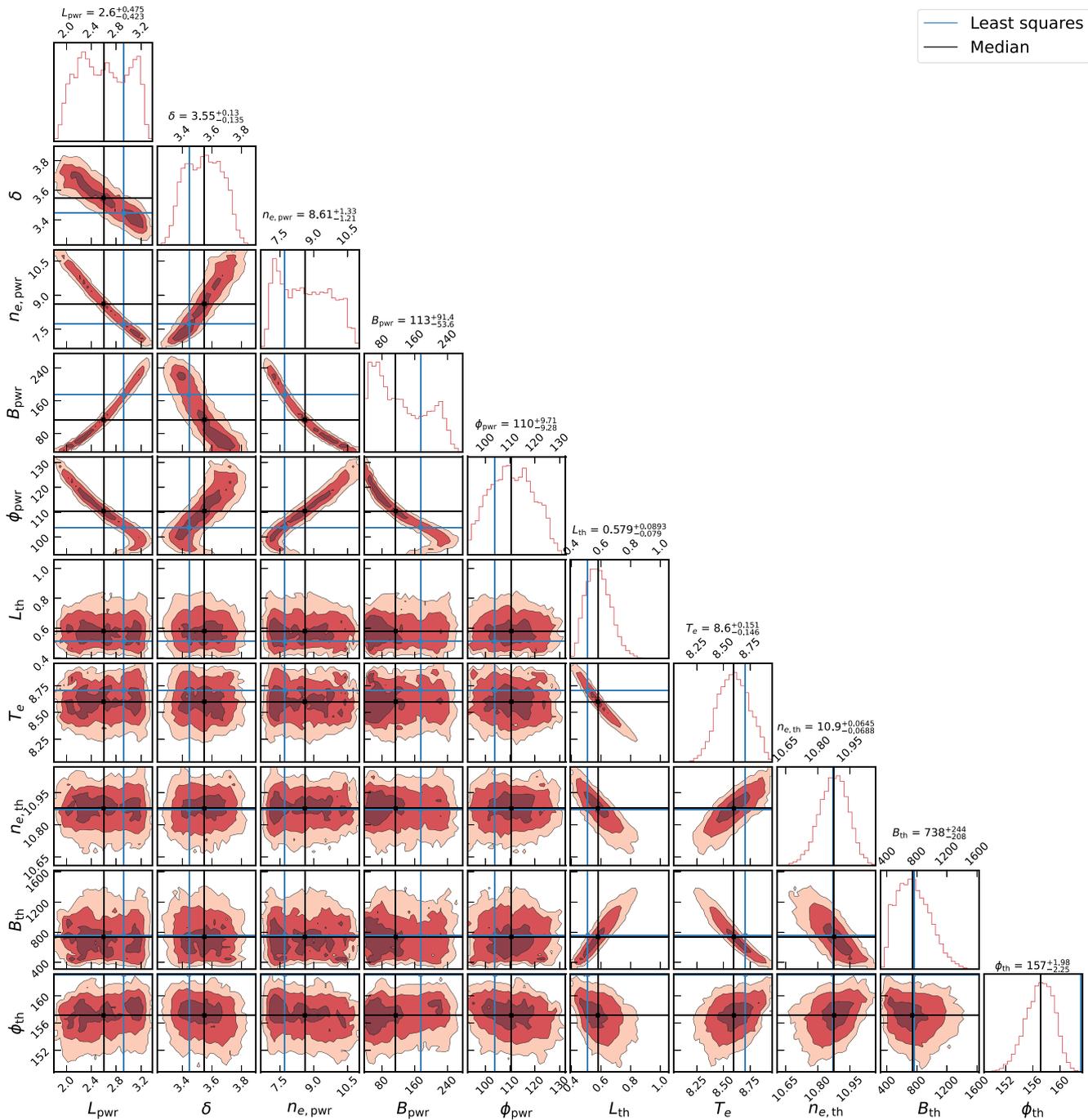

**Figure 8.** Same as Fig. 4 for HD 283572. Note the turnover in the direction of the correlations in $\phi_{\rm pwr}$. This effect is likely due to the approach of a line-of-sight nearly orthogonal to the *B*-field and thus the condition of equation (6) is not met.

## 6 DISCUSSION

The principle observational result of this paper is that thermal GS radiation has been detected for the first time in a non-solar environment, from two WTT stars. This implies that there are regions of very hot ($T_e \sim 10^8$ K) dense thermal plasma with very strong ($\sim 1$ kG) magnetic fields residing somewhere in their stellar magnetospheres. The volumes associated with these thermal GS regions, of order $0.1\,\rm R_\odot^3$, is smaller than the volumes found for the power-law GS emission components, which also have much weaker fields and which are found for all five stars in the sample. The most intriguing, but challenging question is where in the stellar environment is this hot plasma located.

In the following sections, we consider three possible locations: (1) within or near the inner edge of a postulated accretion disc, (2)

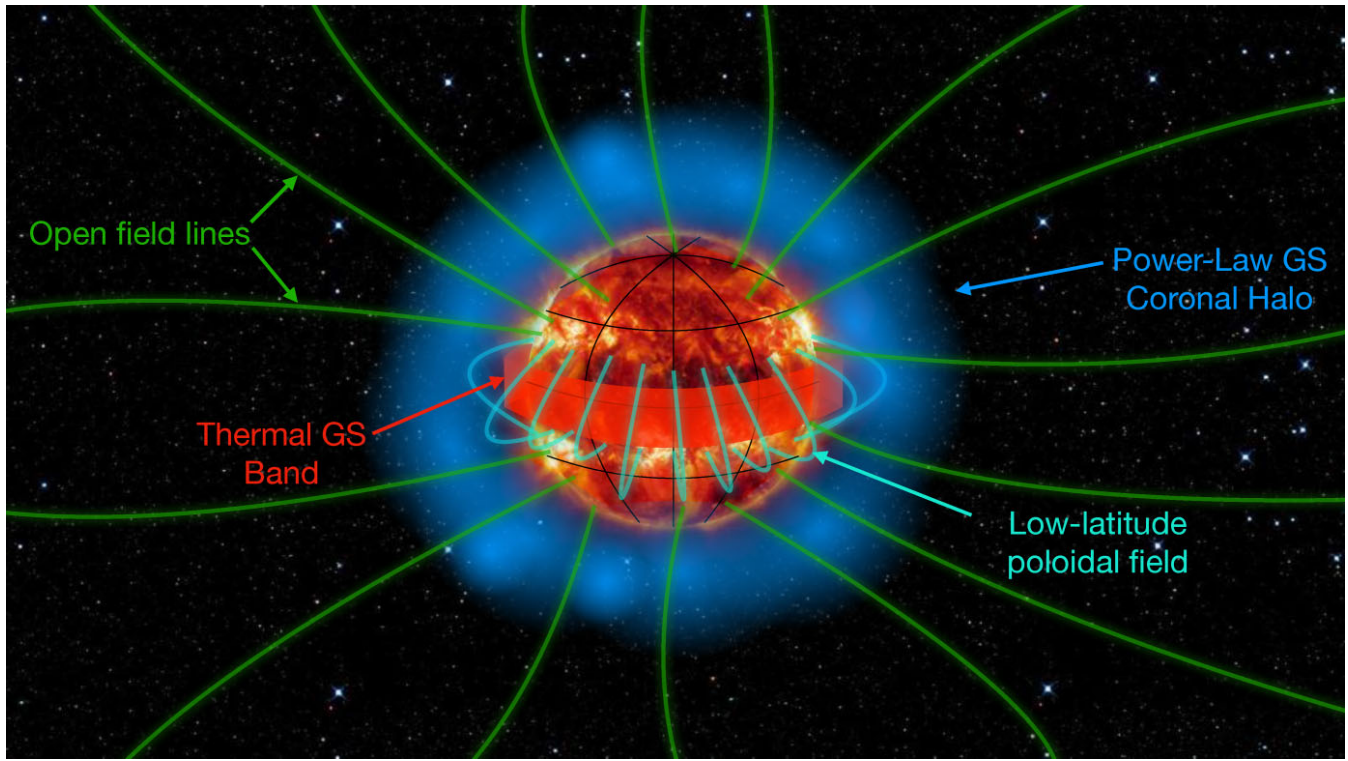

**Figure 9.** Conjectural locations of power-law (blue shading) and thermal (red shading) GS emission regions from the pre-main-sequence objects V410 Tau and HD 283572. The magnetospheric geometry is informed by the ZDI maps of V410 Tau (Finociety et al. 2021) and other WTTS, which show a large low-latitude poloidal component of closed field lines (light green) and numerous open field lines (aqua) extending in all directions.

in an extended magnetosphere with tangled magnetic fields, and (3) within an axisymmetric toroidal field region detected at low magnetic latitudes, as found in WTTs.

### 6.1 TGS emission from a magnetosphere–accretion disc interaction

Classical T Tauri (CTT) stars have moderately strong IR excesses with decreasing fluxes toward longer wavelengths, consistent with emission from opaque dusty circumstellar accretion discs (Hartmann, Herczeg & Calvet 2016). As material from the disc accretes on to the photosphere of the star, the resulting heating creates enhanced chromospheric and coronal emission lines and continua emission detected in the UV and optical bands. If stellar magnetic field lines, which are co-rotating with the star, cannot pass through the inner-disc gas freely, the field lines converging on the inner disc (e.g. Gómez de Castro & Marcos-Arenal 2012, Fig. 15) may twist and shear resulting in reconnection events, consequent particle acceleration, and heating of the ambient plasma. This could be origin of episodic X-ray flares (Romanova et al. 1998) and possibly GS radio emission.

On the other hand, WTTs generally have little or no IR excess and thus no detectable disc emission. Indeed, recent near-IR spectroscopy of both V410 Tau and HD 283 572 provides an upper limit to any putative disc mass: $M < 0.0004\,M_\odot$ (Yasui et al. 2019). This could be taken as strong evidence for these stars having no disc, consistent with the evolutionary paradigm that WTT stars have evolved from the CTT stage after depleting their discs via accretion. However, the observational evidence is not so clear: Many WTT stars show 'composite' time-variable spectra consistent with a disc that has episodic accretion (Littlefair et al. 2004; Gras-Velázquez & Ray 2005; Cieza et al. 2013).

If we conjecture that there may be optically thin circumstellar discs around V410 Tau and HD 283 572 that are currently not accreting, there is the possibility that their stellar magnetospheres may extend to these discs and provide an environment for reconnection, particle acceleration, and plasma heating. A quantitative estimate of the relevant plasma parameters (temperature, density, magnetic field strength) to compare with model parameters derived from the TGS model, would require knowledge of disc properties such as electron density and magnetic diffusivity (e.g. Küker, Henning & Rüdiger 2003), which are unknown. However, assuming the inner disc is of order 5 – 10 stellar radii from the star (Gómez de Castro & Marcos-Arenal 2012), the ambient magnetic field would be diminished by 100x to 1000x at the inner disc, assuming a $r^{-3}$ dipolar radial scaling (higher order fields would diminish faster).

For the case of V410 Tau, whose surface strength is of order 1 kG (Finociety et al. 2021), this implies a field $B \sim 1$–$10\,\mathrm{G}$ at the disc, neglecting reconfiguration due to currents in the disc. These values are strongly at odds with the thermal GS emission model, which requires magnetic fields of order 1 kG. Therefore, even if thin discs were present in these WTT stars, we find this location for either power-law or thermal GS emission untenable.

### 6.2 TGS emission from an extended magnetosphere

We next consider whether TGS emission could arise from an extended stellar magnetosphere. Vidotto et al. (2010) have calculated the magnetic topology for a tilted magnetosphere, wherein the rotation axis and (initially dipolar) magnetic axis are significantly misaligned. They find that the dipolar field is strongly distorted, with larger tilt resulting in greater distortion. If the magnetosphere is filled with sufficiently dense plasma, there is a possibility that the twisted field lines could reconnect, resulting in particle acceleration and ensuing radio emission.

Yu et al. (2019) used ZDI to measure the tilt angle between the rotation axis and the magnetic orientation of V410 Tau's magnetosphere over several years. Although the magnetic topology is complex, they found a dipolar component whose magnitude and orientation is highly variable. The pole is tilted at different angles depending on epoch, from 54° to 18° over the years from 2009 to 2013. This suggests that tangled field line reconnection sites may provide the necessary conditions for either power-law or thermal GS emission.

Estimates of the electron density in the extended coronae of CTT stars are in the range $10^8 - 10^{10}$ cm$^{-3}$ (Colombo et al. 2019), which is in the same range as power-law model densities, and marginally lower than thermal GS models (Table 3). However, as with the interacting disc option, the requisite magnetic field at the emission site seems too low, at least for thermal GS emission. If we use the linear extent $L$ of the power-law model region, $L \sim 1.2 - 2$ R$_\odot$ $\sim 0.4 - 0.7$ stellar radii, a surface magnetic field of order 1 kG, as measured by ZDI, would be of order 100 – 200 G, scaling as $r^{-3}$ as above. This may be compatible with power-law models, which requires $\sim$100 G fields, but not thermal GS models, which require $\sim$ 1 kG fields.

**6.3 TGS emission from low-latitude poloidal fields**

Perhaps the most surprising discovery arising from ZDI mapping of pre-main-sequence stars has been the detection of a strong low-latitude poloidal magnetic component in both WTTs that have been mapped with ZDI so far: LkCA4 (Donati et al. 2019) and V410 Tau (Yu et al. 2019; Finociety et al. 2021). Although the overall magnetic topology is complex and contains both azimuthal and radial components, the ZDI maps show a clear and obvious band of closed coronal loops in meridional planes with heights 0.2 – 0.4 R$_\odot$ girding the magnetic equator (e.g. Yu et al. 2019, Figs. 5, 15). Yu et al. (2019) and Finociety et al. (2021) term this feature 'toroidal', but in this context, they are describing the azimuthal distribution of magnetic field loops rather than the orientation of magnetic field lines.

This is very different from CTT stars, which have relatively simple nearly dipole fields (Donati et al. 2010, 2013), as well as fully-convective M dwarfs more massive than 0.5 M$_\odot$ (Morin et al. 2008). Since WTT stars are usually considered an evolutionary intermediary between CTT pre-main-sequence stars and the main-sequence, the appearance of a completely different magnetic topology is an unexplained puzzle.

The mean magnetic energy of the low-latitude poloidal component for V410 Tau varies from 30 to 50 per cent of the total energy, and has a mean field strength up to 400 G (Yu et al. 2019). Although we have no direct measurements of electron densities or temperatures in this region, it is plausible to assume that conditions in the closed poloidal loops are suitable for particle acceleration via reconnection at the top of the loops, as is well-described in solar flares. Furthermore, the total volume enclosed by the loops is more than required for either the power-law GS model or the thermal GS model.

Since we have previously argued that these regions are distinct, it is reasonable to assign the power-law GS emission region to the extended magnetosphere and the thermal GS region to the denser low-latitude belt. This scheme is illustrated in Fig. 9.

**ACKNOWLEDGEMENTS**

This work made use of data taken by the Karl G. Jansky Very Large Array. The National Radio Astronomy Observatory is a facility of the National Science Foundation operated under cooperative agreement by Associated Universities, Inc. This research made use of the lm-fit non-linear least-squares minimization PYTHON library (Newville et al. 2014), and astropy, a community developed core PYTHON package for astronomy (The Astropy Collaboration 2018). We also made use of EMCEE, an MIT licensed pure-PYTHON implementation of the affine invariant MCMC ensemble sampler (Foreman-Mackey et al. 2013). We also used the PYTHON module corner.py to visualize multidimensional results of MCMC simulations (Foreman-Mackey 2016). This research has made use of the SIMBAD data base, operated at CDS, Strasbourg, France (Wenger et al. 2000). The authors acknowledge Weaver et al. (2021) for the inspiration to provide all code used to generate figures. We thank the anonymous referee for a careful review and helpful comments.

**DATA AVAILABILITY**

We provide Table 2 as our observed Stokes I and V flux densities. Visibilities are available for download from the National Radio Astronomy Observatory Data Archive. The latest version of the code used to generate each figure is hosted on the Github repository for this project. An archival release of the code coincident with publication is also hosted at doi:10.5281/zenodo.7783327.

**REFERENCES**

Bastian T. S., Benz A. O., Gary D. E., 1998, ARA&A, 36, 131
Bauldry S., 2015, in Wright J. D.ed., International Encyclopedia of the Social and Behavioral Sciences (Second Edition), second edition edn, Elsevier, Oxford, p. 615, , https://www.sciencedirect.com/science/article/pii/B9780080970868440559
Carroll T. A., Strassmeier K. G., Rice J. B., Künstler A., 2012, A&A, 548, 95
Cieza L. A., et al., 2013, ApJ, 762, L100
Colombo S., Orlando S., Peres G., Reale F., Argiroffi C., Bonito R., Ibgui L., Stehlé C., 2019, A&A, 624, 50
Donati J. F., 1999, MNRAS, 302, 457
Donati J. F., Semel M., 1990, Solar Physics, 128, 227
Donati J. F., Semel M., Rees D. E., Taylor K., Robinson R. D., 1990, A&A, 232, 1
Donati J. F., Semel M., Rees D. E., 1992, A&A, 265, 669
Donati J. F., Semel M., Carter B. D., Rees D. E., Collier Cameron A., 1997, MNRAS, 291, 658
Donati J. F., et al., 2010, MNRAS, 409, 1347
Donati J. F., et al., 2013, MNRAS, 436, 881
Donati J. F., et al., 2019, MNRAS, 483, 1
Drake S. A., Simon T., Linsky J. L., 1992, ApJS, 82, 311
Dulk G. A., 1985, ARA&A, 23, 169
Dulk G. A., Melrose D. B., White S. M., 1979, ApJ, 234, L1137
Favata F., Micela G., Sciortino S., 1998, A&A, 337, 413
Favata F., Micela G., Reale F., 2001, A&A, 375, 485
Finociety B., et al., 2021, MNRAS, 508, 3427
Fleishman G. D., Kuznetsov A. A., 2010, ApJ, 721, L1127
Foreman-Mackey D., 2016, The Journal of Open Source Software, 1, 24
Foreman-Mackey D., Hogg D. W., Lang D., Goodman J., 2013, PASP, 125, 306
Franciosini E., et al., 2007, A&A, 468, 485
García-Sánchez J., Paredes J. M., Ribó M., 2003, A&A, 403, 613
Gelman A., Gilks W. R., Roberts G. O., 1997, The Annals of Applied Probability, 7, 110
Gelman A., Meng X.-L., Brooks S., Jones G. L., 2011, Handbook of markov chain Monte Carlo. CRC Press (Taylor Francis Group)
Goodman J., Weare J., 2010, Communications in Applied Mathematics and Computational Science, 5, 65
Gordon Y. A., et al., 2021, ApJS, 255, 30
Gras-Velázquez À., Ray T. P., 2005, A&A, 443, 541

Gómez de Castro A. I., Marcos-Arenal P., 2012, ApJ, 749, L190
Hartmann L., Herczeg G., Calvet N., 2016, ARA&A, 54, 135
Karzas W. J., Latter R., 1961, ApJS, 6, 167
Kobayashi K., et al., 2006, ApJ, 648, L1239
Küker M., Henning T., Rüdiger G., 2003, ApJ, 589, L397
Kuznetsov A. A., Fleishman G. D., 2021, ApJ, 922, L103
Lacy M., et al., 2020, PASP, 132, 035001
Launhardt R., Loinard L., Dzib S. A., Forbrich J., Bower G. C., Henning T. K., Mioduszewski A. J., Reffert S., 2022, ApJ, 931, L43
Leto P., Trigilio C., Buemi C. S., Umana G., Leone F., 2006, A&A, 458, 831
Leung P. K., Gammie C. F., Noble S. C., 2011, ApJ, 737, L21
Liddle A. R., 2007, MNRAS, 377, L74
Littlefair S. P., Naylor T., Harries T. J., Retter A., O'Toole S., 2004, MNRAS, 347, 937
Morin J., et al., 2008, MNRAS, 384, 77
Mutel R. L., Molnar L. A., Waltman E. B., Ghigo F. D., 1998, ApJ, 507, L371
Ness J. U., Schmitt J. H. M. M., Burwitz V., Mewe R., Raassen A. J. J., van der Meer R. L. J., Predehl P., Brinkman A. C., 2002, A&A, 394, 911
Newville M., Stensitzki T., Allen D. B., Ingargiola A., 2014, LMFIT: Non-Linear Least-Square Minimization and Curve-Fitting for Python, Zenodo,
Nindos A., Kundu M. R., White S. M., Shibasaki K., Gopalswamy N., 2000, ApJS, 130, 485
Osten R. A., Bastian T. S., 2006, ApJ, 637, L1016
Peterson W. M., Mutel R. L., Güdel M., Goss W. M., 2010, Nature, 463, 207
Peterson W. M., Mutel R. L., Lestrade J. F., Güdel M., Goss W. M., 2011, ApJ, 737, L104
Petrosian V., 1981, ApJ, 251, L727
Phillips R. B., Lonsdale C. J., Feigelson E. D., 1991, ApJ, 382, L261
Pritchard J., et al., 2021, MNRAS, 502, 5438
Retter A., Richards M. T., Wu K., 2005, ApJ, 621, L417
Robinson P. A., Melrose D. B., 1984, Australian Journal of Physics, 37, 675
Romanova M. M., Ustyugova G. V., Koldoba A. V., Chechetkin V. M., Lovelace R. V. E., 1998, ApJ, 500, L703
Schwarz G., 1978, Annals of Statistics, 6, 461
Sokal A. D., 1996, Nuclear Physics B Proceedings Supplements, 47, 172
Storey M. C., Hewitt R. G., 1995, PASA, 12, 174
Sullivan K., Kraus A. L., 2022, ApJ, 928, L134
Tak H., Ghosh S. K., Ellis J. A., 2018, MNRAS, 481, 277
Tan B., 2022, RAA, 22, 072001
Telleschi A., Güdel M., Briggs K. R., Audard M., Scelsi L., 2007, A&A, 468, 443
The Astropy Collaboration, 2018, astropy v3.1: a core python package for astronomy, Zenodo
The CASA Team, et al.,2022, PASP, 134, 1041
Torricelli-Ciamponi G., Franciosini E., Massi M., Neidhoefer J., 1998, A&A, 333, 970
Trigilio C., Buemi C. S., Umana G., Rodonò M., Leto P., Beasley A. J., Pagano I., 2001, A&A, 373, 181
Trubnikov B. A., 1958, PhD thesis, Dissertation, Moscow (US-AEC Tech. Inf. Service, AEC-tr-4073 [1960]), (1958)
Umana G., Trigilio C., Hjellming R. M., Catalano S., Rodono M., 1993, A&A, 267, 126
Vidotto A. A., Opher M., Jatenco-Pereira V., Gombosi T. I., 2010, ApJ, 720, L1262
Walter F. M., et al., 1987, ApJ, 314, L297
Waterfall C. O. G., Browning P. K., Fuller G. A., Gordovskyy M., 2019, MNRAS, 483, 917
Weaver I. C., et al., 2021, AJ, 161, 278
Wenger M., et al., 2000, A&AS, 143, 9
Williams P. K. G., 2018, in Handbook of Exoplanets, Vol. 472, Handbook of Exoplanets. Handbook of Exoplanets, p. 589, , http://link.springer.com/10.1007/978-3-319-55333-7_171
Yasui C., et al., 2019, ApJ, 886, L115
Yu L., et al., 2019, MNRAS, 489, 5556

## APPENDIX A: DEFAULT PRIORS

Careful selection of PDFs toimplement priors is necessary to make statistically sound claims from the posterior (Tak, Ghosh & Ellis 2018). Prior PDFs must be normalizable, even if the explicit normalization is not relevant. The generalized Gaussian and the inverse-gamma function are common options for implementing scientific knowledge in the form of a prior. The generalized Gaussian function is given by

$$p(x) \propto \exp(-|(x-\mu)/\sigma|^s), \tag{A1}$$

where $\mu$ adjusts the centre of the Gaussian, $\sigma$ modifies the width, and $s$ determines how rapidly the fall-off occurs beyond $1\sigma$. Selecting $s = 2$ is a good choice for previous measurements that include uncertainty, that is, a standard Gaussian distribution. Larger $s$ can apply boundaries on parameter space where a model is no longer valid. Implementing a lower-bound is best done via the inverse-gamma function, given by

$$p(x) \propto x^{-a-1}\exp(-b/x), \tag{A2}$$

where a 'soft' lower limit is given by $\frac{b}{a+1}$ and $a$ and $b$ modify the region around the limit, respectively. This PDF is a good option when parameters have sharp lower cutoffs as an alternative to the Heaviside function, which is not normalizable.

We implemented a default prior PDF if a star did not have prior measurements for a particular parameter. These options are summarized in Table A1. The PDFs selected for $\delta$, $\phi$, and $T_e$ are motivated by the validity of the approximate expressions for $\eta_\nu$ and $\kappa_\nu$ in Robinson & Melrose (1984) and Dulk (1985). In the case of

**Table A1.** Default priors.

| Component | Parameter | Unit | Generalized Gaussian | | | Inverse gamma | |
| --- | --- | --- | --- | --- | --- | --- | --- |
| | | | $\mu$ | $\sigma$ | $s$ | $a$ | $b$ |
| Power law | $L$ | [$R_\odot$] | – | – | – | 0.1 | 0.022 |
| | $\delta$ | | 4 | 2.5 | 10 | – | – |
| | $\log(n_e)$ | [cm$^{-3}$] | – | – | – | 1 | 2 |
| | $B$ | [G] | 265 | 265 | 10 | – | – |
| | $\phi_{(1,2)}$ | [deg] | 90 | 70 | 20 | – | – |
| Thermal | $L$ | [$R_\odot$] | – | – | – | 0.1 | 0.022 |
| | $\log(T_e)$[a] | [K] | 8 | 1 | 10 | – | – |
| | $\log(n_e)$ | [cm$^{-3}$] | – | – | – | 1 | 2 |
| | $B$ | [G] | – | – | – | 1 | 2 |
| | $\phi$ | [deg] | 90 | 65 | 20 | – | – |

*Notes.* [a] We implemented the prior on the electron temperature $T_e$ even if there were values of $L_X$ from the literature to guarantee that the electron temperature satisfies the conditions set out by the approximations in Robinson & Melrose (1984).

the *B*-field of the power-law region, the expressions are only valid if $\nu/\nu_B > 10$. If $\nu_B = 2.8B$ MHz and the lowest observed frequency is 15 GHz, then we can solve for a maximum *B* strength of ∼530 G. Finally, the PDFs applied to *L*, $n_e$, and the thermal *B* prevent drifting into non-physical negative values.

## APPENDIX B: 'GOODNESS-OF-CHAIN'

Autocorrelation analysis is a necessary component of any publication employing MCMC sampling. The autocorrelation time is a useful measure of how well an MCMC run has converged by evaluating whether the chain is independent of its initial position. The Monte Carlo standard error can be found from the variance

$$\sigma^2 = \frac{\tau_f}{N}\hat{\sigma}_n,\quad\text{(B1)}$$

where $\tau_f$ is the integrated autocorrelation time for the chain and $\hat{\sigma}_n$ is the variance of the chain. (The familiar ordinary Monte Carlo error of $1/\sqrt{N}$ can be recovered if the chain samples are independent, i.e. $\tau_f = 1$). Thus, an estimation of the autocorrelation time of the chain is a direct measurement of the error.

We stochastically define convergence as if the total length of the chain exceeds 50 times the autocorrelation time. To determine if the chain meets this threshold, the chain is correlated with itself at equal intervals throughout the sampling. MCMC runs that do not meet this threshold have the number of steps increased proportionally. We point the reader to Sokal (1996) and Gelman et al. (2011) for a derivation of Monte Carlo standard error and further discussion of evaluating chain independence.

We may also qualitatively evaluate a chain by plotting the 1D density of each parameter of each walker versus the step number, called a 'walker plot'. An MCMC run that has converged will demonstrate relatively constant density over the same region for the entire chain length. Large-scale trends in the movement of the walkers or diverging densities indicate instability in posterior space. Fig. B1 shows an example of a successful MCMC walker plot from the burn-in steps of the run for UX Arietis.

The acceptance fraction quantifies how often a walker 'accepts' a step. It takes a step either due to a preferable value of the posterior or against the posterior gradient. This will occur with a probability proportional to the ratio of the posterior values at the current and proposed location. For optimal speed in characterizing the posterior probability distribution, the acceptance fraction should approach 0.23 (Gelman et al. 1997). For an MCMC chain that implements the 'stretch-step' algorithm, $\alpha$ is a number greater than 1 that parametrizes the possible step lengths, where a walker will step in the direction of another randomly selected walker with uniform probability of lengths between $1/\sqrt{\alpha}$ and $\sqrt{\alpha}$ (Goodman & Weare 2010). Reducing $\alpha$ can generally increase the acceptance fraction (and therefore the average step length), but the autocorrelation time is likely to grow, so a modification of the chain's total length may also be necessary.

The final settings and results of the EMCEE runs used in this work are reported in Table B1. In all cases, we ran chains until the autocorrelation condition was met and the acceptance fraction fell between 0.1 and 0.5.

**Table B1.** EMCEE settings and results.

| Star | $\alpha$ | Burn-in | Production | Autocorrelation | Acceptance fraction |
|---|---|---|---|---|---|
| HR 1099 | $\frac{3}{2}$ | 75 000 | 1 500 000 | 11 900 (126x) | 0.27 |
| UX Arietis | 2 | 2000 | 20 000 | 210 (95x) | 0.46 |
| Algol | $\frac{5}{4}$ | 150 000 | 2 500 000 | 31 900 (79x) | 0.25 |
| V410 Tau | $\frac{4}{3}$ | 37 500 | 750 000 | 7350 (102x) | 0.25 |
| HD 283572 | $\frac{5}{4}$ | 100 000 | 2 000 000 | 24 400 (81x) | 0.21 |

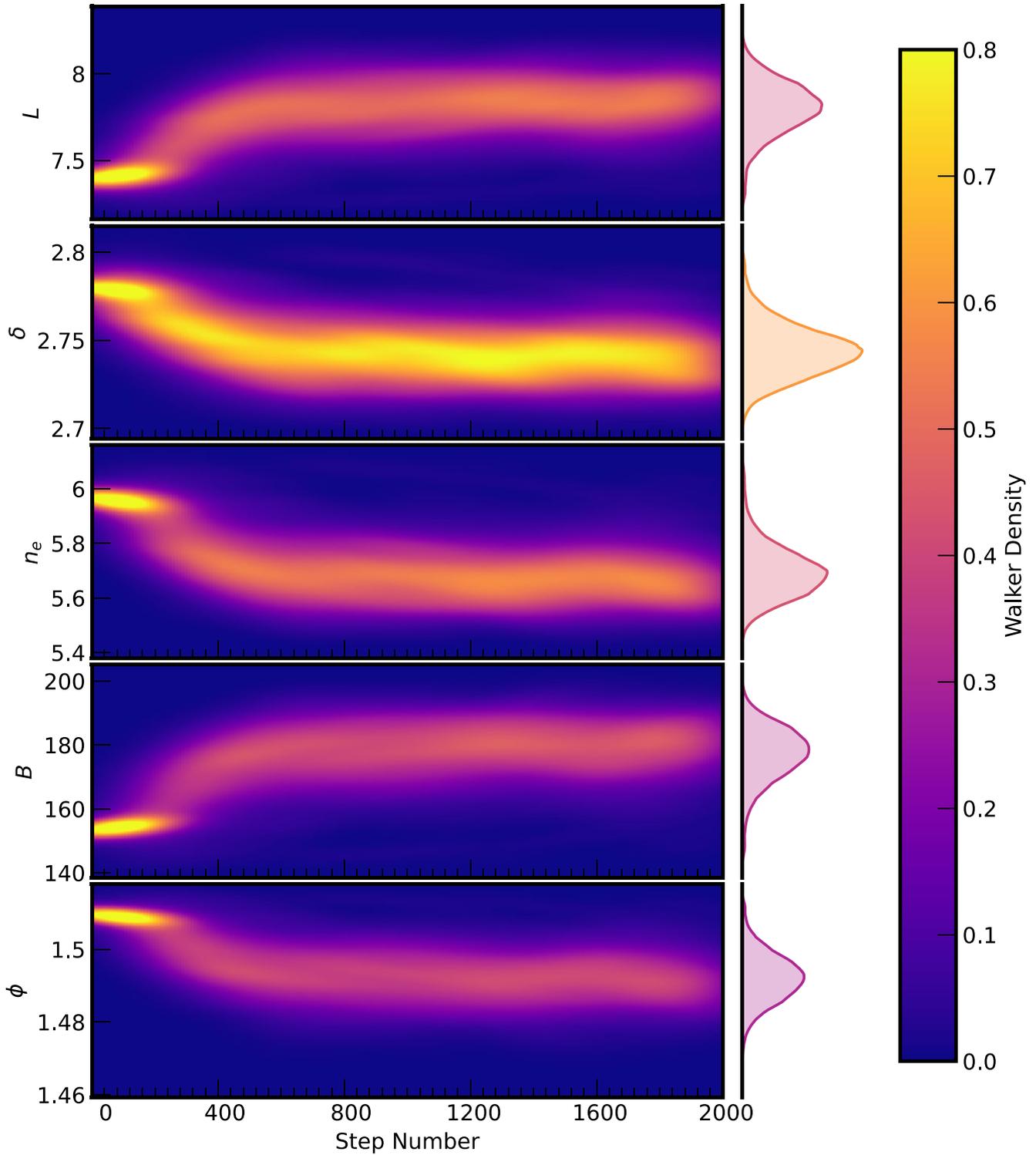

**Figure B1.** Walker position in each parameter as a function of step number. Convergence of the walkers is obvious by consistent and repeated exploration of the same region of posterior space. The distributions on the right side of each plot indicate the median value of the walker density along that axis. The distribution is shaded according to the maximum of the median values.

This paper has been typeset from a T<sub>E</sub>X/LAT<sub>E</sub>X file prepared by the author.